\documentstyle[aps,prl,epsf]{revtex}

\newcommand{\be}{\begin{equation}}
\newcommand{\ee}{\end{equation}}
\newcommand{\ba}{\begin{eqnarray}}
\newcommand{\ea}{\end{eqnarray}}

\begin{document}

\draft

\twocolumn[\hsize\textwidth\columnwidth\hsize\csname@twocolumnfalse%
\endcsname

\title{Doping dependence of the N\'eel temperature in Mott-Hubbard
antiferromagnets: Effect of vortices}
\author{Carsten Timm and K.H. Bennemann}
\address{Institut f\"ur Theoretische Physik, Freie Universit\"at
Berlin, Arnimallee 14, D-14195 Berlin, Germany}
\date{\today}

\maketitle

\begin{abstract}
The rapid destruction of long-range antiferromagnetic order upon doping
of Mott-Hubbard antiferromagnetic insulators is studied within a
generalized Berezinskii-Kosterlitz-Thouless renormalization group
theory in accordance with recent calculations suggesting
that holes dress with vortices.
We calculate the doping-dependent N\'eel temperature in
good agreement with experiments for high-$T_c$ cuprates.
Interestingly, the critical
doping where long-range order vanishes at zero temperature is
predicted to be $x_c\sim 0.02$, independently of any energy scales
of the system.
\end{abstract}

]

\narrowtext

%% INTRODUCTION

The study of lightly doped Mott-Hubbard (MH) antiferromagnetic insulators
is of great current interest, since the insulating parent compounds of
cuprate high-$T_c$ superconductors are of this type. The various
parts of the phase diagram of these compounds are believed to be intimately
related. Therefore it is important to understand the properties of
the antiferromagnetic
phase, in particular the rapid destruction of magnetic order upon doping and
the anomalously small critical doping of $x_c\sim 0.02$
holes per copper ion \cite{Luke}. In the present letter we
derive the N\'eel temperature $T_N$ as a function of doping using a
generalized Berezinskii-Kosterlitz-Thouless (BKT) renormalization group
theory \cite{BKT,Minn} for the vortices in the antiferromagnetic state.
There are two types of vortices, thermally created electrically neutral ones
and electrically charged ones, which are centered at the holes. Both types are
nucleated separately (as vortex-antivortex pairs), but additively screen
the vortex interaction, with a common unbinding temperature $T_N(x)$.
This temperature is indeed strongly reduced upon doping
and vanishes at $x_c\sim 0.02$
independently of the energy scales of the system.
Our approach is independent of any particular microscopic model and can
thus serve as a guide for electronic theories.

\begin{figure}[ht]
\centerline{\epsfxsize 2.50in\epsfbox{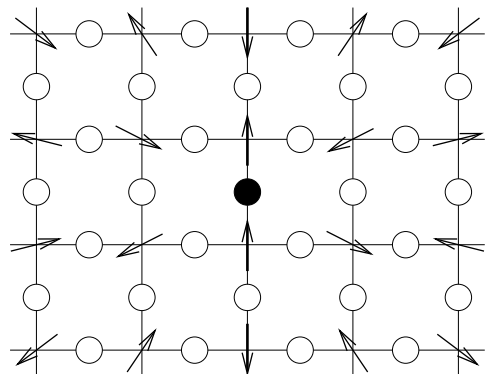}}
\caption{Schematic representation of a
charged vortex induced by a single hole at an
oxygen site (black circle). The hole frustrates the interaction
between copper spins (arrows). The circles denote oxygen atoms.
A neutral vortex would be centered on the square between
four oxygen atoms and not have
a hole in its core.}
\label{fig1}
\end{figure}

Our physical picture is the following: The holes introduced by doping
are mainly located at the planar oxygen sites,
where they frustrate the antiferromagnetic exchange interaction
between copper spins due to their tendency to
form copper-oxygen spin singlets \cite{Emery,SBB}. This may
lead to ferromagnetic coupling between the two spins \cite{ferro}.
Since the system is approximately two-dimensional,
the staggered magnetization can form a vortex as sketched
in Fig.~\ref{fig1} to evenly distribute the frustration induced by the hole.
On the other hand, neutral vortices without a hole in their core
can be thermally created. Due to the easy-plane
Dzyaloshinskii-Moriya anisotropy \cite{DM},
the staggered magnetization can be described
by a two-component order parameter at low energies, leading to
a logarithmic size dependence of the single-vortex energy \cite{Verges,BJ}.
This implies a logarithmic vortex interaction, making BKT scaling ideas
applicable.
To describe the interplay of charged and neutral vortices, which
determines the N\'eel temperature $T_N$,
we have to extend the BKT theory to a system with two kinds of topological
defects. One important point is that
the density of charged vortices is given by the
doping $x$. The other is that we have to describe the screening of
the vortex interaction due to both types.

Electronic theory supports our picture:
Within unrestricted Hartree-Fock theory
Verg\'es {\it et al.}~\cite{Verges} found several competing low-energy
configurations for a lightly hole-doped Hubbard model, including spin
polarons, domain walls, and holes dressed with vortices and antivortices.
In a spin polaron the staggered local moments in the vicinity
of the hole are reduced but still collinear, while in a vortex (antivortex)
they rotate through $2\pi$ ($-2\pi$).
Seibold \cite{GS} using a slave-boson approach and
Berciu and John \cite{BJ} within a self-consistent Hartree-Fock
theory found that an even number of holes
dress with vortices and antivortices (or merons \cite{BJ})
in the ground state for appropriate parameters.
Since the energy of a single vortex diverges (logarithmically
\cite{Minn,Verges,BJ}) with
system size, whereas that of a vortex-antivortex pair remains
finite, only pairs are created in infinite systems.

An advantage of BKT-type theories is that they include
fluctuations on all length scales, in particular on large ones, which
are crucial close to the phase transition.
Previous studies that only included fluctuating local moments without
correlations between them overestimated
the critical doping $x_c$ \cite{ZAB}. Similarly, the decrease of $T_N$
with $x$ obtained from the fluctuation exchange approximation
is also too slow \cite{FLEX}. By including correlations between
neighboring spins into a slave-boson theory for the three-band
Hubbard model, Schmalian {\it et al.}\ \cite{SBB}
obtained a critical doping of $x_c\sim 0.025$ in better agreement
with experiment. However, this approach takes only fluctuations on
the length scale of the lattice constant into account, but
neglects fluctuations on larger length scales.

%% THEORY

In BKT theory the interaction of vortices
is screened by the polarization of vortex pairs lying between them
\cite{BKT,Minn}. As the temperature is increased more pairs
are thermally created leading to increased screening. At the
N\'eel temperature the screening becomes strong enough
for the largest pairs to break up. The resulting free vortices
destroy the magnetic order.
The situation here is more complex: First, we have to take account
of the screening due to both
charged and neutral vortices, and second, the density of charged
vortices is fixed by the doping level $x$.

Only neutral pairs are thermally
created, whereas charged vortices enter the system only upon doping.
In principle one could imagine a single hole doped into the system to form
a pair consisting of a charged vortex and a neutral antivortex or {\it
vice versa}, but 
microscopic calculations \cite{Verges,GS,BJ} do not find this configuration
at $T=0$. Rather, two holes are needed to produce a vortex-antivortex pair.
For simplicity we assume this to hold also at finite temperatures.
However, even if mixed pairs are not created upon doping, they are formed
when vortex pairs exchange partners.

The density of neutral vortices is controlled by
their chemical potential $\mu_{\text{neu}}$ or equivalently
the vortex core energy $E_{\text{core}}=-\mu_{\text{neu}}$, which
depends on details of the copper-oxygen and
copper-copper interactions and is treated here as a parameter.
The energy of a vortex-antivortex pair is $2E_{\text{core}}+V$
with the interaction \cite{NFL} $V(r) = q^2\,\ln (r/r_0)$,
where $q^2=2\pi J S^2$ is the strength of the interaction,
$J$ is the exchange interaction between nearest neighbors,
$S=1/2$ is the spin, and $r_0$ is the small-distance cutoff
of BKT theory. $r_0$ can be interpreted as the smallest possible
vortex-antivortex separation \cite{BKT,Minn}.
Two charged vortices additionally experience a Coulomb
interaction, which, however, is irrelevant in the renormalization-group
sense, since it falls off faster than $V(r)\sim \ln r$.

The probability of creating a neutral or charged vortex in
an area $r_0^2$ is given by
its fugacity $y_{\text{neu}}$ and $y_{\text{ch}}$, respectively.
Since we have assumed that vortices are only created in neutral or
charged pairs,
we consider the pair fugacities $y^2_{\text{neu}}$ and $y^2_{\text{ch}}$.
For the smallest possible neutral pairs of size $r_0$,
\be
y^2_{\text{neu}}(r_0) = C^2_{\text{neu}} e^{2\beta\mu_{\text{neu}}} ,
\label{ini1}
\ee
where $C_{\text{neu}}$ is a constant of order unity \cite{BKT} and
$\beta$ is the inverse temperature.
The constraint on the density of charged vortices is implemented by
choosing $y^2_{\text{ch}}(r_0)$ in such a way that their total density
equals the hole density, see below.
The vortex interaction is screened by the
polarization of smaller vortex pairs,
$V(r) = q^2/\epsilon(r)\:\ln(r/r_0)$. The screening is described by the
spin-wave stiffness $K(r) = \beta q^2/2\pi\epsilon(r)$.
In the renormalization group, small pairs of sizes between $r$ and $r+dr$
are integrated out and their effect is approximately incorporated
into renormalized quantities $K(r)$, $y^2_{\text{neu}}(r)$, and
$y^2_{\text{ch}}(r)$. Starting from $r=r_0$,
this operation is repeated for larger and larger
pairs leading to the recursion relations \cite{BKT,TGF}
\ba
dy_\eta^2/dl & = & 2(2-\pi K)\, y^2_\eta ,
\label{K1} \\
dK/dl & = & -4\pi^3 (y^2_{\text{neu}}+y^2_{\text{ch}}) K^2 .
\label{K2}
\ea
The initial conditions are Eq.~(\ref{ini1}) and $K(r_0)=\beta q^2/2\pi$.
Equation (\ref{K1}) determines the fugacities of neutral
($\eta=\text{neu}$) and charged ($\eta=\text{ch}$) pairs of size
$r=r_0 e^l$. Two separate equations for $y_{\text{neu}}^2$ and
$y_{\text{ch}}^2$ are present, since we assume that
vortices are created either as neutral pairs or as
charged pairs with two holes. Both types feel the same
screened interaction $V$ at large distances so that the same stiffness
$K$ appears. Differences
at smaller separation are incorporated into the core energies.
Equation (\ref{K2}) describes the additional screening due to pairs of
size $r_0 e^l$. Their total density is proportional to
$y^2_{\text{neu}}+y^2_{\text{ch}}$.

If the stiffness $K$ vanishes for $l\to\infty$, the interaction
is fully screened for large pairs ($\epsilon\to\infty$),
which thus become unbound, destroying the magnetic order.
Since the interaction on large length scales is the same for
neutral and charged vortices, this unbinding happens at a single
transition for both types.
While solving Eqs.~(\ref{K1}) and (\ref{K2}) we have to
simultaneously satisfy the constraint on the density $n_{\text{ch}}$
of charged vortex pairs. As shown in Ref.~\cite{CT},
this density can be expressed in terms of the fugacity,
\be
n_{\text{ch}} = \int_{r_0}^\infty \!\! dr\,2\pi r\,\frac{y^2_{\text{ch}}(r)}
  {r^4} = \frac{2\pi}{r_0^2} \int_0^\infty \!\!
  dl\,e^{-2l}\,y^2_{\text{ch}}(l) .
\ee
The pair density has to equal half the
density of holes, $n_{\text{ch}} = x/2a^2$,
where $a$ is the lattice constant.
In practice, the recursion relations are integrated numerically to find
$y^2_{\text{ch}}(l)$, from which we calculate $n_{\text{ch}}$. The initial
value $y^2_{\text{ch}}(0)$ is varied until the contraint is satisfied.

\begin{figure}[ht]
\centerline{\epsfxsize 3.00in\epsfbox{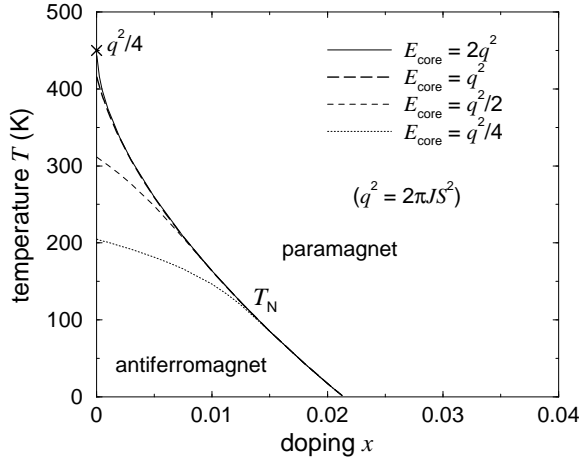}}
\caption{Phase diagram for the antiferromagnetic phase of the cuprates.
The N\'eel temperature $T_N$ is calculated using
four values of the core energy $E_{\text{core}}$
of thermally created vortices. Note
that the curves for larger $E_{\text{core}}$ are very close to
the one for $E_{\text{core}}=2q^2$. The symbol $\times$ at zero
doping denotes the maximal possible transition temperature.}
\label{fig2}
\end{figure}

The resulting phase diagram is shown in Fig.~\ref{fig2}. We used
$C_{\text{neu}}=1$, $J=1800\text{ K}$, and
$r_0=2a$, and varied the core energy $E_{\text{core}}$.
The phase below the N\'eel temperature $T_N$ shows quasi-long-range
antiferromagnetic order, which is made long range by the weak
interlayer exchange.
The phase for $T>T_N$ or $x>x_c$ is characterized by free vortices, which
destroy the long-range order, but leave short-range order on the length
scale of the separation between free vortices intact. Short range
correlations have indeed been observed in cuprates up to much larger
dopings. For $T\to 0$ we find the critical pair density,
\be
n^c_{\text{ch}} \approx 0.04273\: r_0^{-2}.
\ee
The numerical factor is {\it universal\/}: Since
neutral vortices do not exist for $T\to 0$, it cannot
depend on $E_{\text{core}}$ and $C_{\text{neu}}$.
The remaining energy scale $q^2$ does not enter the result, since it
is multiplied by the diverging $\beta=1/k_BT$.
While $n^c_{\text{ch}}$ and thus
the critical doping $x_c=2\,n^c_{\text{ch}}a^2$ are independent of
energy scales, they do depend on
the {\it non-universal\/} minimal separation $r_0$ of two vortices
which is of the order of twice the core radius or twice
the correlation length. Slave-boson and Hartree-Fock calculations
\cite{GS,BJ} show that the core radius
is not significantly larger than a lattice spacing $a$.
In Fig.~\ref{fig2} we have taken the core radius to be
$a$ so that $r_0=2a$, which results in $x_c\sim 0.021$
in very good agreement with experiments on
${\mathrm La}_{2-x}{\mathrm Sr}_x{\mathrm CuO}_4$
\cite{Luke,Keimer,Borsa,muSR}. For
${\mathrm Y}_{1-z}{\mathrm Ca}_z{\mathrm Ba_2Cu_3O_6}$ experiments
find $z_c/2\sim 0.03$ \cite{muSR}, of the same
order of magnitude as our value of $0.021$ (the number of holes per
copper atom is $z/2$ in this double-layer compound).
For ${\mathrm YBa_2Cu_3O}_{7-\delta}$ a critical hole concentration
of $0.021$ corresponds to $\delta_c\sim 0.68$, following
Tallon {\it et al.} \cite{Tallon}, in good agreement with
experiments \cite{Burns}.

In the high-temperature region of the phase diagram
the overall temperature scale is set by the maximal possible N\'eel
temperature $T_N^{\text{max}}=q^2/4=\pi J S^2/2$. For $S=1/2$ this gives
$T_N^{\text{max}}\approx 0.393\,J$, compared to the mean-field
result $T_N^{\text{mf}} = 0.5\,J$ for a Heisenberg
antiferromagnet on a cubic lattice with interlayer exchange $J_\perp\ll J$.
The reduction of $T_N$ is due to fluctuations,
which are strong for quasi-two-dimensional systems. The actual
value of $T_N(x=0)$ and the shape of the curve $T_N(x)$ at small
doping are determined by the core energy $E_{\text{core}}$.
Note, we obtain the correct temperature scale under
the reasonable assumption that $E_{\text{core}}$
is not very much smaller than the interaction strength $q^2$.
Experimentally, $T_N$ is found to depend only
weakly on doping for small $x$ \cite{Borsa}, which requires a small core
energy. Then many neutral vortices are created
at a given temperature so that charged vortices only become
relevant at higher doping. Conversely, for
large core energy only a few thermal vortices are present even
at $T_N^{\text{max}}$. For $E_{\text{core}} \gtrsim 2q^2$ the curve
$T_N(x)$ in Fig.~\ref{fig2} does not change
appreciably with $E_{\text{core}}$ so that
the charged vortices would determine the magnetic properties even
at very small doping.

\begin{figure}[ht]
\centerline{\epsfxsize 3.00in\epsfbox{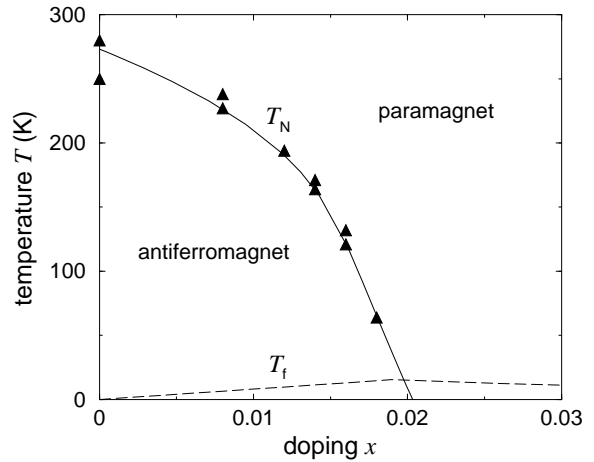}}
\caption{Comparison of the calculated N\'eel temperature for
$E_{\text{core}}/q^2\approx 0.13$ (solid line) and results for $T_N$
from NQR and $\mu$SR experiments for
${\mathrm La}_{2-x}{\mathrm Sr}_x{\mathrm CuO}_4$
(triangles) \protect\cite{Borsa}. The dashed line schematically shows
the freezing temperature $T_f$ \protect\cite{muSR}, below which the holes
become immobile.}
\label{fig3}
\end{figure}

Quantitative agreement with $\mu$SR and NQR experiments 
on ${\mathrm La}_{2-x}{\mathrm Sr}_x{\mathrm CuO}_4$ by
Borsa {\it et al}.\ \cite{Borsa}
can be obtained by appropriate choices of the exchange $J$,
the core energy $E_{\text{core}}$, and the core radius $r_0$, see
Fig.~\ref{fig3}. For this plot, $J=2410\mbox{ K}$,
$E_{\text{core}}=0.1303 q^2=493\mbox{ K}$,
and $r_0=2.052 a$. Typical experimental values are
$J\sim 1400\mbox{ K}$ \cite{JLa}.
This discrepancy may be due to the simplified description of the anisotropic
antiferromagnet by two-component spins or to the neglect of
the interlayer exchange $J_\perp$ and the doping dependence of $J$.

We now briefly comment on electron-doped cuprates. While we reproduce
the correct critical doping for hole-doped compounds, our approach
would not yield the much larger critical doping $x_{c,e}\sim 0.14$
in electron-doped cuprates \cite{Luke}. The reason is that
the additional electrons mainly
fill up the copper $3d$ orbitals and destroy the magnetic moments
at the copper sites. Thus, the main effect of electron doping is to
dilute the antiferromagnet. There is no spin-singlet formation
involved and hence no tendency towards vortex formation. Then differently
as in the case of hole doping $T_N$ decreases due to spin dilution.

The validity of our approach is questionable if the charged
vortices become immobile. There is
experimental evidence that this happens below a crossover
temperature $T_f$, which increases with doping and reaches about
$16\mbox{ K}$ in ${\mathrm La}_{2-x}{\mathrm Sr}_x{\mathrm CuO}_4$
for $x\approx x_c$
and falls off again for larger $x$ \cite{GSM,muSR}.
$T_f$ is sketched in Fig.~\ref{fig3}.
Below this temperature the holes (charged vortices)
form a glass and their dynamics strongly slows down.
The experimental N\'eel temperature in this region
should depend strongly on the time scale of the
experiment. On the other hand, the true phase transition is governed
by the behavior in the limit of infinite time
so that the formation of a glass below $T_f$ should affect it
only weakly. The doping dependence of $T_f$ can be qualitatively
understood in our picture:
In the magnetically disordered phase the logarithmic part of
the interaction of charged vortices
is screened on the length scale of the correlation length
\cite{BKT,Minn,CT}, which is still large close to $T_N$.
Thus the interaction of charged vortices (holes)
changes smoothly at the transition and decreases
for larger doping, leading to similar behavior of the freezing
temperature $T_f$.

There are theoretical indications that the holes may form
one-dimensional stripes at low temperatures \cite{WS}.
Modeling stripes by a phenomenological anisotropic Hei\-sen\-berg
model, Castro Neto and Hone \cite{CNH} calculated the
N\'eel temperature $T_N(x)$ within a renormalization-group scheme
and found good agreement with experiment. However, in this theory
the critical concentration $x_c$ is basically a free parameter.
Note, stripes formed by charged vortices consist of
alternating vortices and antivortices, in order to
lower the interaction energy. These stripes are
automatically anti-phase domain walls \cite{JLTP}, which are observed
experimentally.

%% CONCLUSIONS

In conclusion, by starting from the assumption that holes doped into
the Mott-Hubbard antiferromagnet dress with vortices and using
independently obtained values
for the exchange interaction and the antiferromagnetic correlation length
in the ordered phase, we obtain a doping-temperature
phase diagram for the antiferromagnetic phase in qualitative agreement with
experiment. In particular, the predictions for the critical doping at zero
temperature and the N\'eel temperature at zero doping are of the observed
order of magnitude. Our approach uses a generalized
BKT theory, which does not depend on any particular microscopic model.
With an appropriate choice of the core energy of
thermally created vortices we can obtain quantitative agreement with
experiment. The core energy controls
the shape of the phase boundary at small
doping, but does not affect the region of higher doping,
where the N\'eel temperature approaches zero at a critical hole density
that is universal in natural units.
Our results show that stripes are
not required to understand the data. The success of this
theory based on vortex fluctuations emphasizes the importance
of two-dimensionality for understanding the cuprates. This
should also hold in the more strongly doped superconducting
region. It would be desirable to
include the spin rearrangement around holes, which is induced by
frustration due to singlet formation, into an electronic theory
of underdoped cuprates. For the antiferromagnetic phase such a
theory should yield results similar to the ones shown here.

\end{document}